\documentstyle[epsfig]{lamuphys}
\makeatletter
\let\chapter\hid@chapter
\makeatother
\begin{document}
\pagenumbering{arabic}
\title{Nucleon-Nucleon Interaction and Isospin Violation
\thanks{Invited plenary talk at Workshop on Chiral Dynamics 1997,
Mainz, Sept 1-5, 1997}}

\author{U. van Kolck \thanks{Address after Jan 1 1998: 
                            W.K. Kellogg Radiation Laboratory, Caltech, 
                            Pasadena, CA 91125}}

\institute{Department of Physics, University of Washington, Seattle, 
           WA 98195-1560, USA}

\maketitle

\begin{abstract}
The application of the 
chiral effective theory to processes with two or more 
nucleons is discussed. 
We gain a qualitative understanding of the gross features of nuclear physics
and quantitative, testable postdictions and predictions
involving photons and pions. 
\end{abstract}
\section{Introduction}
\newcommand{\lsim}{\mathrel{\rlap{\lower4pt\hbox{\hskip1pt$\sim$}}\raise1pt\hbox{$<$}}}
\newcommand{\pronabla}{\vec{\nabla}}
\newcommand{\boldpi}{\stackrel{\rightarrow}{\pi}}
\newcommand{\boldtau}{\stackrel{\rightarrow}{\tau}}
\newcommand{\boldT}{\raise1pt\hbox{$\stackrel{\rightarrow}{T}$}}
\newcommand{\boldD}{\raise1pt\hbox{$\stackrel{\rightarrow}{D}$}}
\newcommand{\boldt}{\stackrel{\rightarrow}{t}}
\newcommand{\boldepsilon}{\stackrel{\rightarrow}{\epsilon}}
 
My topic is not yet part of mainstream
chiral perturbation theory ($\chi$PT): 
the low-energy 
effective field theory (EFT) for systems with more than one nucleon. 
This subject is fascinating because it not only involves the
symmetries of QCD, but also demands  an understanding 
of a non-trivial interplay between non-perturbative and perturbative
physics.

I will try to emphasize here the aspects of the problem that are not 
standard in other applications of $\chi$PT. 
In particular, my main goal is to make sense of nuclear physics, 
rather than search for tests of chiral dynamics. (Although, as you will see, 
we are also opening a window on a whole new set of tests of 
$\chi$PT.) 
By ``to make sense'' I mean to formulate the problem in 
terms of an EFT, so that after the relevant degrees of freedom and symmetries 
are determined we can devise
an expansion in powers of $Q/M$, where $Q$ represents the
typical momentum of the processes we are interested in
and $M$ stands for a characteristic mass scale of the underlying 
theory.
If this is accomplished, then we will have made nuclear physics rooted in QCD 
(consistent with chiral symmetry), systematic (amenable to a perturbation 
treatment) and applicable to all processes where $Q\ll M_{QCD}\sim 1$ GeV
(a theory, rather than a model).

The questions I would like to address are: 
Why are  nuclei so loosely bound
compared to $M_{QCD}$?
Why are processes involving nucleons and external probes
(pions and photons) generically dominated by two-body interactions? 
Why is isospin a very good symmetry at low energies? 
In the first part of the talk I concentrate on  
shallow bound states in EFTs, a problem that can be discussed without 
explicit pions. In the second part I briefly review some of the 
results that have been achieved in few-nucleon problems where pions do 
play a very important role. In the third part isospin violation is discussed.

\section{Very low energies}
In processes with typical momenta 
$Q$ much smaller than the pion mass $m_\pi$,  
the only relevant degree of freedom is the nucleon of mass $m_N$,
the important symmetries are parity, time-reversal and ``small''
Lorentz boosts, 
and the appropriate expansion parameter is 
$Q/M \sim \partial / (m_\pi, m_N, ...)$.
(Electromagnetic processes can also be considered by adding the photon,
$U(1)_{em}$ gauge invariance, and $\alpha_{em}$ to this list.) 
Much effort has been spent during the last year in trying to understand 
issues related to regularization and fine-tuning of 
this ``pionless'' theory 
(\cite{vkolck:kaplanetal}, \cite{vkolck:kaplan}, \cite{vkolck:cohen},
\cite{vkolck:phillips1}, \cite{vkolck:scaldeferri},
\cite{vkolck:lukeman}, 
\cite{vkolck:lepage}, 
\cite{vkolck:richardsonetal},
\cite{vkolck:phillips2}, \cite{vkolck:beaneag},
\cite{vkolck:vkolck1}, \cite{vkolck:bedaque}).

The most general Lagrangian with such input consists of 
an infinite number of contact terms, which are quadratic, quartic, ...,  
in the nucleon fields $N$ with increasing number of derivatives: 
\begin{eqnarray}
\cal L & = & N^\dagger \left(i\partial_{0}
                           +\frac{\pronabla^2}{2m_N}
                           +\ldots\right) N
           - \frac{C_0}{2} N^\dagger N N^\dagger N     \nonumber \\
  &  & + \frac{C_2}{8} 
         \left[N^\dagger \pronabla N \cdot N^\dagger \pronabla N 
          - N^\dagger N N^\dagger\pronabla^2 N 
          - N^\dagger N \pronabla^2 (N^\dagger N) + \mbox{h.c.}\right] \nonumber \\
  &  & +\ldots,                       \label{E:vkolck:lag}
\end{eqnarray}
\noindent
where $C_{2n}$ are parameters of mass dimension $-2(n+1)$.
Here, to avoid cluttering the discussion with 
unnecessary detail, I omitted spin and isospin combinations, 
the $Q^4/m_N^3$ relativistic correction, and
a two-derivative four-nucleon interaction that contribute to
$P$-waves, and lumped terms that only contribute to higher orders in 
``$\ldots$''.
Nucleons are non-relativistic 
and 
the corresponding field theory has nucleon number conservation.

Let me first consider the two-nucleon system in the center-of-mass frame,
with total energy denoted by $k^2/m_N$.
The two-nucleon amplitude $T_{NN}$ 
is simply a sum of bubble graphs, 
whose vertices are the four-nucleon contact terms that appear 
in the Lagrangian (\ref{E:vkolck:lag}). 
It consists of 
two different expansions, a loop expansion and an expansion 
in the number of insertions of derivatives at the vertices or nucleon lines.
Compare for example the one-loop graph 
made out of two $C_0$ vertices to the tree-level graph from $C_0$.
Their ratio is
\begin{equation}
m_N C_0 \int \frac{d^3l}{(2 \pi)^3}\frac{1}{l^2-k^2 -i\epsilon}
= \frac{m_N C_0}{4\pi}[\theta \Lambda+ ik+ O(k^2/\Lambda)] 
\label{E:vkolck:int}
\end{equation}
\noindent
where I introduced a regulator $\Lambda$. Both the number $\theta$ 
and the function $O(k^2/\Lambda)$ depend on the regularization scheme 
chosen. This dependence is mitigated by renormalization.
The linear divergence can be absorbed in $C_0$ itself, 
$1/C_0^{(R)}=1/C_0 +m_N\theta\Lambda/4\pi+\ldots$, while
to correctly account for the $k^2$ term, at least one insertion of $C_2$ is
necessary. 
The loop expansion is therefore an expansion in $m_N Q C_0^{(R)}/4\pi$, while
the derivative expansion is in $C_2^{(R)} Q^2/C_0^{(R)}$ 
and similar combinations of 
the higher order coefficients.

In a natural theory, there is no fine-tuning and one scale $M$. In such 
case we expect all the parameters to scale with $M$. 
If 
$C_{2n}^{(R)}= 4\pi \gamma_{2n}/m_N M^{2n+1}$ with 
$\gamma$'s all dimensionless
parameters of $O(1)$, then 
the loop expansion is in $Q/M$ and the derivative expansion in
$(Q/M)^2$.
$T_{NN}$ is perturbative and 
equivalent to an effective range expansion for 
$k\ll 1/a \sim 1/\sqrt{a r_0}\sim \ldots \sim M$,
where
the $S$-wave scattering length is $a= m_N C_0^{(R)}/4\pi$,
the $S$-wave effective range is 
$r_0 = 16\pi C_2^{(R)}/m_N (C_0^{(R)})^2$, and so on.
This scaling of effective range parameters is indeed what one gets in 
simple potential models, like a square well of range $R \sim 1/M$ and
depth $V_0 \sim M$.
Similarly, one can show that in a wave of angular momentum $l$,
the relevant mass scale in the expansion is $(2l+1) M$.
Since it is perturbative, the amplitude in this EFT can only describe 
scattering, not bound states.

In nuclear physics, we are interested in 
shallow bound states. This means that the underlying theory has at least 
one parameter $\alpha$ which is fine-tuned to be close to a critical 
value $\alpha_c$ at which there is a bound state 
at zero energy.
In this case there exist {\it two} distinct scales: 
$M$ and  $\aleph = (\alpha-\alpha_c) M \ll M$.
Assume that 
$C_{2n}^{(R)}= 4\pi \gamma_{2n}/m_N \aleph (M \aleph)^n$ 
with 
$\gamma$'s again all dimensionless parameters of $O(1)$.
(This can be accomplished with natural-size bare coefficients 
if they are fine-tuned against a regulator $\Lambda \sim M$; 
for example, 
$C_0=-(4\pi/ \theta m_N \Lambda) (1+ \aleph/\gamma_0 \theta \Lambda
+\ldots)$.)
In this case the loop expansion is in 
$Q/\aleph$, while the derivative expansion is in
$Q^2/\aleph M$.
We are justified in resumming the larger $Q/\aleph$ terms,
which can be done analytically.
This produces a leading order amplitude
$-C_0^{(R)}/(1+im_N C_0^{(R)} k/4\pi)$.
With the loop expansion so resummed, the derivative expansion becomes an 
expansion in $C_2^{(R)} Q^2/ C_0^{(R)} (1+im_N C_0^{(R)}Q/4\pi)$
and similar combinations. That is, an expansion in
$Q^2/M(\aleph +iQ)$, which fails at momenta $\sim 1/M$.
For $Q\ll \aleph$ the theory is still perturbative, but a (virtual or real) 
bound state appears at 
$k= i (4\pi/ m_N C_0^{(R)}) (1 + ...)\sim (\pm) i \aleph$.
The bound state lies in the 
region of applicability of the EFT,
which can thus be used in a non-trivial way.

For momenta $Q\sim \aleph$, the dominant sub-leading terms 
are terms in $C_2^{(R)}$ which start at $O(\aleph/M)$.
Interactions that generate higher effective range parameters 
(such as the shape parameter)
only contribute at $O((\aleph/M)^3)$ or higher. 
It simplifies calculations considerably to also resum the $C_2^{(R)}$
interaction,
which is easily done for again they are in a  geometric series.
Such a resummation generates a four-fermion interaction of the type
$C_0^{(R)}/(1-2(C_2^{(R)}/C_0^{(R)}) k^2)$. 
It was noted by \cite{vkolck:kaplan} that
this resembles the exchange of an s-channel particle, 
provided 
{\it (i)} the sign $\sigma$ of its kinetic term 
and its coupling to nucleons $g^{(R)}$
satisfy 
$\sigma (g^{(R)})^2 =- (C_0^{(R)})^2/2m_N C_2^{(R)} \sim \pm 2 \pi M/m_N^2$;
and 
{\it (ii)} its mass is
$\Delta^{(R)}= C_0^{(R)}/2m_N C_2^{(R)} \sim \pm M \aleph/2m_N$.
An alternative EFT can then be constructed 
by writing the most general Lagrangian consistent with the same symmetries 
as above, but now including an additional
light di-baryon field for each (real or virtual) bound state.
Because the symmetries are the same as the EFT with nucleons only,
the two EFTs give the same result at any given order in the derivative
expansion. 
The EFT with a di-baryon field has  bare poles
(ghosts with negative kinetic energy for $r_0 \geq 0$)
which change character upon dressing:
one becomes the
(real or virtual) bound state at $k \sim \pm i/\aleph$,
the other being a ghost at $k\sim i/M$, outside the range of the EFT.
Since in $NN$ scattering there are two $S$-wave bound states near threshold,
one introduces two di-baryon fields, an $S=1, I=0$ $\vec{D}$ and
an $S=0, I=1$ $\boldT$, and in leading order
there are two masses and two couplings which 
are fitted to the triplet and (average) singlet scattering lengths and 
effective ranges.  

Further resummations can be done similarly, and they can likewise
be reproduced by other alternative EFTs with more di-baryon fields that can 
mix. I will not pursue this here. 
It is clear that to any given order the EFT for short-range
interactions is equivalent to an effective range expansion to the same 
order: 
\begin{equation}
T_{NN}(k) =  
- \left( \frac{1}{C_0^{(R)}} -2 \frac{C_2^{(R)}}{(C_0^{(R)})^2} k^2 
         +\frac{im_N k}{4\pi}+ O(Q^4/\aleph M)\right)^{-1}, \label{renTon}
\end{equation}
\noindent
which gives $a \sim \aleph^{-1}$, while other effective range parameters
still scale with $M$. 
Once more, this is in agreement with potential model results,
for example a square well with $\sqrt{m_N V_0} R$ close to $\pi/2$.
From the measured $NN$ scattering lengths, we see that
fine-tuning is considerable in the $^1S_0$ channel
($^1\aleph \sim 10$ MeV), but less severe in the 
$^3S_1$ channel ($^3\aleph \sim 40$ MeV).

The same result (\ref{renTon})
can be obtained more directly by first summing
insertions of $C_2$ to all orders and then expanding in $Q^2$.
If the higher orders are kept, regularization scheme dependence will remain
through the $Q^4$ and higher terms.
However, this dependence will be no greater than 
the dependence
on neglected higher order interactions if $\Lambda \sim M$. 
Formally, this is equivalent to solving a Schr\"odinger 
equation with a 
bare potential consisting, schematically, of a sum 
$C_0 \delta_{\Lambda}(\vec{r}) + C_2 \delta_{\Lambda}''(\vec{r}) +...$, 
where $\delta_{\Lambda}$ is a regulated delta function.
Effectively, we replace the ``true'', possibly complicated potential
of range $\sim 1/M$ by a multipole expansion with moments $C_{2n}$. 
It is not difficult to show (\cite{vkolck:vkolck1}) that the
effect of renormalization is to turn this bare potential into a 
generalized pseudo-potential
$(C_0^{(R)}+ 2 C_2^{(R)} k^2 +...) \delta(\vec{r})
\frac{\partial}{\partial r} r$, 
or equivalently, turn the problem into a free 
one with boundary conditions at the origin which are analytic in the energy. 
The first, energy-independent term, 
parametrized by $C_0^{(R)}$,
was considered by \cite{vkolck:bethe}.

I now turn to an example of 
how the application of these ideas to other nuclear systems can bring
non-trivial model-independent predictions. (Sometimes called low-energy
theorems.)

Consider the three-nucleon system. At momenta $Q\sim \aleph$, only 
scattering states are accessible, so I confine myself to 
nucleon-deuteron scattering. For illustration, take the zero energy case, 
where $S$-waves dominate. There are two channels, a quartet of total spin
$J=3/2$ and a doublet of $J=1/2$. 
The leading interactions 
involve only two-body interactions conveniently written 
via the di-baryon fields:
assuming naive dimensional analysis, 
three-nucleon forces start only at $O((\aleph/M)^4)$.
The $Nd$ scattering amplitude 
is the sum of these two-body interactions to all orders; 
in the quartet only $\vec{D}$ contributes 
while in the doublet $\boldT$ also appears.
This results in a particularly simple set of Faddeev equations:
there is one integral equation in one variable in the quartet
and a pair of coupled integral equations in the doublet channel.
The quartet $Nd$ scattering length can then be obtained 
in a model independent way with {\it no free parameters}. 
Solving the integral equation,
\cite{vkolck:bedaque} 
found $^4a=6.33$ fm with an uncertainty from higher orders 
of $\sim \pm 0.10$ fm. This result is 
in very good agreement with
the experimental value of 
$^4a=6.35\pm 0.02$ fm (\cite{vkolck:dilg}).
Work on the energy dependence of the quartet amplitude and 
on the doublet channel is in progress.

In the same vein, we could consider other $Q\ll m_\pi$ processes.
Unfortunately, this leaves out most of the interesting aspects of nuclear 
physics.
Let me use $2 m_N B/ A$ as a measure of a typical momentum $Q$ of 
a nucleon 
in a nucleus with $A$ nucleons and binding energy $B$. 
(Other quantities such as charge radii give similar estimates.) 
$Q/m_\pi$ is then about 0.3 for $^2$H, 0.5 for $^3$H, 0.8 for $^4$He, ..., 
and 1.2 for symmetric nuclear matter in equilibrium. 
The same argument that justified the use of a pionless 
theory for the deuteron
now suggests that understanding the binding of typical nuclei ($^4$He 
and heavier) requires {\it explicit} inclusion of pions, 
but {\it not} of heavier mesons such as the rho. 
Enter $\chi$PT.

\section{Chiral Effective Theory}

The EFT for typical nuclear momenta 
$Q\ll M_{QCD}$ can be formulated 
along the same lines as the pionless case. 
The extra degrees of freedom 
---besides non-relativistic nucleons and photons---
are obviously pions and also non-relativistic delta isobars, 
since both the pion mass $m_\pi$ and 
the delta-nucleon mass difference $m_{\Delta}-m_N\sim 2m_\pi$ 
are of the order of the momenta $Q$ we want to consider. 
The new and very important symmetry is the approximate 
chiral symmetry $SU(2)_L\times SU(2)_R \sim SO(4)$,
assumed to be broken spontaneously down to its diagonal 
$SU(2)_{L+R} \sim SO(3)$. 
The expansion parameter
is expected to be 
$Q/M_{QCD}\sim (\partial, m_\pi, m_{\Delta}-m_N)/
              (m_\rho, m_N, 4\pi f_\pi, ...)$ 
---besides $\alpha_{em}$.

Were it not for the approximate
chiral symmetry, this would be a very difficult EFT to 
handle, because it would lack any small parameters.
If the quark masses were zero (``chiral limit''), 
the QCD Lagrangian would be invariant under 
independent rotations of the quarks' 
left- and right-handed components,
while the hadronic spectrum suggests only
one isospin rotation is realized.
Goldstone's theorem assures us there is in the spectrum
a scalar boson, the pion $\boldpi$, 
which corresponds to excitations 
in the coset space $SO(4)/SO(3) \sim S^3$. 
We call the radius of this ``circle'' $f_\pi \simeq 92$ MeV.
Since an infinitesimal chiral transformation is of the form 
$\boldpi\rightarrow\boldpi \!\! + f_\pi \!\!\!\boldepsilon+...$,
the Lagrangian will be
$SO(4)$ symmetric if it depends on $\boldpi$ only through 
covariant derivatives, that is, derivatives
on the circle, which are non-linear: 
$\boldD_\mu=(\partial_\mu \!\!\!\boldpi\!\!\!/ 2 f_\pi) (1+ O(\pi^2))$,
${\cal D_\mu} N= (\partial_\mu + 
i \!\!\!\boldt \!\!\!\cdot \!\!\!\boldpi \!\!\!\times \boldD_\mu/f_\pi)N$,
etc. 
Quark masses generate two terms in the QCD Lagrangian. 
One term, $\bar{m} \bar{q}q$ with $\bar{m}=(m_u+m_d)/2$, is the
fourth component of an $SO(4)$ vector and therefore breaks $SO(4)$ 
explicitly down to $SO(3)$ of isospin. 
The effective Lagrangian will acquire then 
an infinite set of terms that do depend on $\boldpi$ in an isospin 
invariant way and transform 
as 4-components of (tensor products of) $SO(4)$ vectors, and have 
coefficients proportional to (powers of) 
the small parameter 
$\eta=\bar{m}/\Lambda_{QCD} \sim m_\pi^2/ \Lambda_{QCD}^2$. 
(Unless the last proportionality factor is zero as argued by
\cite{vkolck:stern}, a possibility I will not entertain here.)
The other quark mass term, $\epsilon \bar{m} \bar{q}\tau_3 q$ with 
$\epsilon=(m_u-m_d)/(m_u+m_d)\simeq 1/3$, is the third component of 
another $SO(4)$ vector and further breaks $SO(3)$ down to $U(1)\times U(1)$. 
This, plus the effects of electromagnetism, will be discussed in 
Section 4.

The most general Lagrangian 
consists of adding pions and deltas to Eq. (\ref{E:vkolck:lag}),
according to the rules above. (See, for example, 
\cite{vkolck:gasser}, \cite{vkolck:bernard}, and \cite{vkolck:kambor}.)
The important point is that these new interactions are weak at low energies 
because of their derivative nature and/or because of pion mass
and delta-nucleon mass difference factors.
We can naturally group the interactions in sets ${\cal L}_{(\Delta)}$,
\begin{equation}
{\cal L} 
 = \sum_{\Delta=0}^{\infty} {\cal L}_{(\Delta)},  \label{E:vkolck:L}
\end{equation}
\noindent
of common index $\Delta\equiv d+q+n+\frac{f}{2} -2$, 
where $d$, $q$, and $n$ are, respectively,
the number of derivatives, powers of $m_\Delta -m_N$, 
and powers of $m_\pi$, and $f$ is the 
the number of fermion fields.
For non-electromagnetic interactions, we find that 
$\Delta\ge 0$ only because of the constraint of chiral symmetry. 
(Electromagnetic contributions can have negative $\Delta$, but this is 
compensated by enough powers of $\alpha_{em}$.)

Consider now an arbitrary irreducible contribution to a process involving
$A$ nucleons and any number of pions and photons, all with momenta 
of order $Q$. 
It can be represented by a Feynman diagram with $A$ continuous nucleon lines,
$L$ loops, $C$ separately connected pieces, and 
$V_\Delta$ vertices from ${\cal L}_{(\Delta)}$, 
whose connected pieces cannot be all split by cutting only 
initial or final lines.
($C=1$ for $A=0,1$; $C=1, ..., A-1$ for $A\ge 2$. The reason to consider
irreducible diagrams and $C> 1$ will be mentioned shortly.) 
It is easy to show (\cite{vkolck:weinberg1}, \cite{vkolck:weinberg2}) 
that this
contribution is typically of $O((Q/M_{QCD})^\nu)$, where
\begin{equation}
\nu=4-A+2(L-C)+\sum_\Delta V_\Delta \Delta.  \label{E:vkolck:nu}
\end{equation}
Since $L$ is bounded from below (0) and $C$ from above ($C_{max}$), 
the chiral symmetry constraint $\Delta\ge 0$
implies that $\nu\ge \nu_{min}=4-A-2C_{max}$ for strong interactions. 
Leading contributions come from tree diagrams built out of ${\cal L}_{(0)}$
and coincide with current algebra. Perturbation theory in $Q/M_{QCD}$
can be carried out by considering contributions from ever increasing $\nu$.

In the case of strong mesonic processes, 
$\nu=2+2L+\sum_\Delta V_\Delta \Delta\geq 2$
with $\Delta= d+n-2$ increasing in steps of two. 
For processes where one nucleon is present,    
$\nu=1+2L+\sum_\Delta V_\Delta \Delta \ge 1$, where
$\Delta= d+q+n+\frac{f}{2} -2$ (with $f=0, 2$) increases in steps of one.
In both cases, all there is is perturbation theory.
Thanks to this power counting, a low-energy nucleon can in a very 
definite sense be pictured as a static, point-like object 
(up to corrections in powers of $Q/m_N$), surrounded by: 
{\it (i)} an inner cloud which is dense but of short range $\sim 1/m_\rho$, 
so that we can expand in its relative size $Q/m_\rho$;
{\it (ii)} an outer cloud of long range $\sim 1/m_\pi$ but sparse,
so that we can expand in its relative strength $Q/4 \pi f_\pi$.

As we have seen in Section 2,
a non-trivial new element enters the theory when we consider systems of
more than one nucleon (\cite{vkolck:weinberg1}, \cite{vkolck:weinberg2}).
Because nucleons are heavy, contributions from intermediate states
that differ from the initial state only in the energy of nucleons are
enhanced by infrared quasi-divergences:
as it can be seen in Eq. (\ref{E:vkolck:int}),
small energy denominators of $O(Q^2/m_N)$ generate contributions
$O(m_N/Q)$ larger than what would be expected from Eq. (\ref{E:vkolck:nu}).
The latter is still correct for the class of sub-diagrams
---called irreducible--- that do not contain intermediate states with
small energy denominators. 
For an $A$-nucleon system these are $A$-nucleon irreducible diagrams, 
the sum of which we call the potential $V$.
When we consider external probes all with $Q\sim m_\pi$, 
the sum of irreducible diagrams forms the kernel $K$ 
to which all external particles are attached.
A generic diagram contributing to a full amplitude will consist
of irreducible diagrams sewed together by states of small energy denominators. 
These irreducible diagrams might have more than one connected piece, 
hence the introduction of $C$ in Eq. (\ref{E:vkolck:nu}).
One way to deal with
the infrared enhancement 
is to introduce a regulator 
$\Lambda \sim M_{QCD}$ and 
sum irreducible diagrams to all orders
in the amplitude, thus creating the possibility of the existence 
of shallow bound states (nuclei). For an $A$-nucleon system, 
this is equivalent to solving the 
Schr\"odinger equation with the bare potential $V$. 
The amplitude for a process
with external probes is then $T\sim \langle \psi'|K|\psi\rangle$ 
where $|\psi\rangle$ ($|\psi'\rangle$) 
is the wavefunction of the initial (final) nuclear state calculated 
with the potential $V$. 
All bare parameters depend on $\Lambda$ but after these are fit to data,
scheme dependence is no
greater than higher orders terms in the Lagrangian.
Because our $Q/M_{QCD}$ expansion is still valid 
for the potential and the kernel,
the picture of a nucleon as a mostly static object surrounded by an inner
and an outer cloud leads to remarkable nuclear physics properties that 
we are used to, but would otherwise remain unexplained from the 
viewpoint of QCD.

If we assemble a few non-relativistic nucleons together, each nucleon will not
be able to distinguish details of the others' inner clouds. The region of 
the potential associated with distances of $O(1/m_{\rho})$ can be expanded in
delta-functions and their derivatives as Bethe and Peierls did. The outer 
cloud of range $O(1/m_{\pi})$ yields non-analytic contributions
to the potential, but being sparse, 
it mostly produces the exchange of one pion, 
with progressively smaller two-, three-, ...- pion exchange contributions. 

\begin{figure}[htb]
\centerline{\epsfig{file=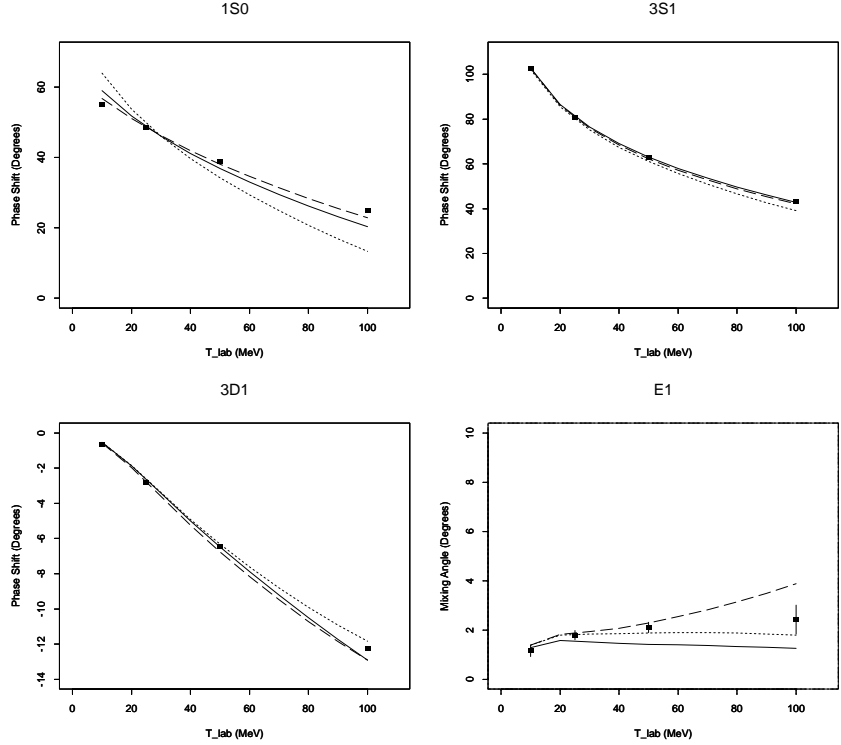,height=6.0in,width=4.5in}}
\vspace{-5.0cm}
\caption{$^1S_0$, $^3S_1$, and $^3D_1$ $NN$ phase shifts and 
$\epsilon_1$ mixing angle in degrees as functions
of the laboratory energy in MeV: 
chiral expansion up to $\nu=3$ for cut-offs of 
500 (dotted), 780 (dashed), and 1000 MeV (solid line); 
and Nijmegen phase shift analysis (squares).}
\label{F:vkolck:NNfigs}
\end{figure}

For the two-nucleon system, 
$\nu=2L+\sum_\Delta V_\Delta \Delta$, with $\Delta$ as in the 
one-nucleon case (but $f=0,2,4$). 
A calculation of all the contributions up to $\nu=3$ was carried out
by \cite{vkolck:ordonez1}, \cite{vkolck:ordonez2}, 
and \cite{vkolck:ordonez3}. 
In leading order, $\nu=0$, the potential is simply static one-pion exchange 
and momentum-independent contact terms (\cite{vkolck:weinberg1}). 
$\nu=1$ corrections vanish due to parity and time-reversal invariance. 
$\nu=2$ corrections include several two-pion exchange diagrams 
(including virtual delta isobar contributions), recoil in one-pion exchange, 
and several contact terms that are quadratic in momenta. 
At $\nu=3$ a few more two-pion exchange diagrams contribute.
As in the pionless case, regularization and renormalization are necessary. 
It is not straightforward to implement dimensional regularization in this 
non-perturbative context, so we used an overall gaussian cut-off, 
and performed calculations with the cut-off parameter
$\Lambda$ taking values 500, 780 and 1000 MeV. 
Cut-off independence means that for each cut-off value a set of 
bare parameters can be found that fits low-energy data. 
As mentioned before, in an $l$ partial wave the onset of non-perturbative
effects is at $(2l+1) m_\pi$;
in high partial waves contact interactions of only high 
order contribute, so these waves are mostly determined
by perturbative pion exchange.
A sample of the results for the lower, more interesting partial waves
is compared to the Nijmegen phase shift analysis (\cite{vkolck:nijmegen})
in Fig. \ref{F:vkolck:NNfigs}, 
and deuteron quantities are presented in 
Table \ref{T:vkolck:dparam}. 
\cite{vkolck:ordonez3} have more details and experimental references.

\begin{table}[htb]
\caption{Effective chiral Lagrangian fits for various cut-offs $\Lambda$
and experimental values 
for the deuteron binding energy ($B$), magnetic moment ($\mu_d$), electric 
quadrupole moment ($Q_E$), asymptotic $d/s$ ratio ($\eta$), and $d$-state 
probability ($P_D$).} 
\label{T:vkolck:dparam}
\begin{tabular*}{\textwidth}{@{}l@{\extracolsep{\fill}}cccc}
\hline
Deuteron   & \multicolumn{3}{c}{$\Lambda$ (MeV)} &            \\
\cline{2-4}
quantities & 500 & 780 & 1000                   & Experiment \\
\hline
$B$ (MeV)  & 2.15 & 2.24 & 2.18                 & 2.224579(9) \\
$\mu_d$ ($\mu_N$) & 0.863 & 0.863 & 0.866       & 0.857406(1) \\
$Q_E$ (fm$^2$) & 0.246 & 0.249 & 0.237          & 0.2859(3)   \\
$\eta$     & 0.0229 & 0.0244 & 0.0230           & 0.0271(4)  \\
$P_D$ (\%) & 2.98 & 2.86 & 2.40                 &            \\
\hline
\end{tabular*}
\end{table}

This is, in Aron Bernstein's terms, a Cadillac calculation,
but in part because it is to be eventually replaced by more economic and
faster machines...
The fair agreement of this first calculation and data 
up to laboratory energies of 100 MeV or so suggests that this may 
become an alternative to other, more model-dependent approaches 
to the two-nucleon problem.
Further examination of 
aspects of two-pion exchange in this context can be 
found in 
\cite{vkolck:celenza}, \cite{vkolck:friar}, \cite{vkolck:carocha1},
\cite{vkolck:carocha2}, \cite{vkolck:carocha3}, \cite{vkolck:carocha4},
\cite{vkolck:savage}, and \cite{vkolck:kaiseretal}.

Obviously, in this EFT theory, too,
we can simultaneously 
get some insight into other aspects of nuclear forces. 
Let us look for the new forces that appear in systems with more than two 
nucleons. 
The dominant potential, at $\nu=6-3A=\nu_{min}$, 
is the two-nucleon potential of lowest order that appeared in 
the two-nucleon case.
We can easily verify that a three-body potential will arise at 
$\nu=\nu_{min}+2$, a four-body potential at $\nu=\nu_{min}+4$, and so on. 
It is (approximate) chiral symmetry therefore that
implies that $n$-nucleon forces $V_{nN}$ are expected to obey a hierarchy
of the type
$\langle V_{(n+1)N}\rangle/\langle V_{nN}\rangle
\sim O((Q/M_{QCD})^2)$,
with $\langle V_{nN}\rangle$ denoting the contribution per $n$-plet.
If we estimate 
$\langle V_{2N}\rangle \sim \frac{g_A^2}{16\pi f_\pi^2} m_\pi^3 \simeq 10$ MeV,
we can guess 
$\langle V_{3N}\rangle \sim 0.5$ MeV,
$\langle V_{4N}\rangle \sim 0.02$ MeV, and so on. 
This is in accord with detailed few-nucleon calculations using more 
phenomenological potentials.
The explicit three-body potential at $\nu=\nu_{min}+2$ (from the delta 
isobar) and $\nu_{min}+3$ was derived by \cite{vkolck:vkolck2}.
It is dominated by the delta, and bears some resemblance
to other, more phenomenological potentials insofar as
two-pion exchange is concerned, but shorter-range pieces are new.

Despite these successful fits and insights, 
the main advantage of the method of EFT lies 
in its concomitant application to many other processes, which
might yield more predictive statements. I now discuss some of these.

As a result of the factor $-2C$ in Eq. (\ref{E:vkolck:nu}), we
see immediately ---in an effect similar to few-nucleon forces--- that
external low-energy probes ($\pi$'s, $\gamma$'s) 
with $Q\sim m_\pi$ will tend to interact 
predominantly with a single nucleon, 
simultaneous interactions with more than 
one nucleon being suppressed by powers of $(Q/M_{QCD})^2$. Again, this 
generic dominance of the impulse approximation is a 
well-known result that arises naturally here. 
This is of course what allows extraction, to a certain accuracy, of  
one-nucleon parameters from nuclear experiments. 
More interesting from the nuclear dynamics perspective are, however,
those processes where the leading single-nucleon contribution vanishes by 
a particular choice of experimental conditions, 
for example the threshold region. 
In this case the two-nucleon contributions, especially in the 
relatively large deuteron, can become important.

{\it $\bullet \hspace{.2cm} \pi d\rightarrow \pi d$ at threshold.}
This is perhaps the most direct way to check the consistency of $\chi$PT in
few-nucleon systems and in pion-nucleon scattering.
Here the lowest-order, $\nu=-2$ contributions to the kernel vanish because 
the pion is in an $S$-wave and the target is isoscalar.
The $\nu=-1$ term comes from the (small) isoscalar pion-nucleon seagull,
related in lowest-order to the pion-nucleon isoscalar amplitude $b_0$.
$\nu=0, +1$ contributions come from corrections to pion-nucleon scattering
and two-nucleon diagrams, which involve besides $b_0$ also the much larger 
isovector amplitude $b_1$. \cite{vkolck:weinberg3} has estimated
these various contributions to the pion-deuteron scattering length, finding 
agreement with previous, more phenomenological calculations,
which have been used to extract $b_0$. 
(See also \cite{vkolck:beanenew}.)

{\it $\bullet \hspace{.2cm} n p \rightarrow \gamma d$ at threshold.}
This offers a chance of a precise postdiction.
Here it is the transverse nature of the real outgoing photon that is 
responsible for the vanishing of the lowest-order, $\nu=-2$ contribution
to the kernel. 
The single-nucleon magnetic contributions come 
at $\nu=-1$ (tree level), $\nu=+1$ (one loop), etc. 
The first two-nucleon term is an one-pion exchange at $\nu=0$,
long discovered to give a smaller but non-negligible contribution.
There has been a longstanding discrepancy of a few percent between these
contributions and experiment. 
At $\nu=+2$ there are further one-pion exchange, two-pion exchange,
and short-range terms. 
\cite{vkolck:park1} and \cite{vkolck:park2}
calculated the two-pion exchange
diagrams in a `` deltaless'' 
theory and used resonance saturation to estimate 
the other $\nu=+2$ terms. With wavefunctions from the Argonne V18 
potential and a cut-off $\Lambda=1000$ MeV,
they found the excellent agreement with experiment shown in 
Table \ref{T:vkolck:radcap}. 
The total cross-section changes by 0.3\% 
if the cut-off is decreased to 500 MeV. 
(See \cite{vkolck:park2} where references to experiment can be found.)

\begin{table}[t]
\caption{Values for various contributions to 
the total cross-section $\sigma$ in mb
for radiative neutron-proton capture:
impulse approximation to $\nu=2$ (imp),
impulse plus two-nucleon diagrams at $\nu=0$ (imp+tn0),
impulse plus two-nucleon diagrams up to $\nu=2$ (imp+tn),
and experiment (expt).} 
\label{T:vkolck:radcap}
\begin{tabular*}{\textwidth}{@{}l@{\extracolsep{\fill}}cccr}
\hline
imp & imp+tn0 & imp+tn & expt\\
\hline
305.6 &  321.7 & 336.0  &  334$\pm$0.5\\
\hline
\end{tabular*}
\end{table}

{\it $\bullet \hspace{.2cm} \gamma d\rightarrow \pi^0 d$ at threshold.}
This reaction offers the possibility to test a
prediction arising from a combination of two-nucleon contributions 
and the neutral pion single-neutron amplitude. 
Pion 
photoproduction on the nucleon has been studied
up to $\nu=4$ in the deltaless theory;
see \cite{vkolck:bernard}. 
A fit constrained by resonance saturation 
is successful in reproducing the measured differential 
cross-section at threshold $\propto |E_{0+}|^2$
in the channels
$\gamma p\rightarrow \pi^+ n$,
$\gamma n\rightarrow \pi^- p$, and
$\gamma p\rightarrow \pi^0 p$, and makes a prediction
$E_{0+}(\gamma n\rightarrow \pi^0 n)=2.13 \cdot 10^{-3}/m_{\pi^+}$. 
Neglecting isospin violation, the uncertainties in
the three measured amplitudes
do not allow to pinpoint $E_{0+}(\pi^0 n)$ much better than
$-0.5\mapsto +2.5 \cdot 10^{-3}/m_{\pi^+}$.
The lack of neutron targets highlights the advantage of using
the deuteron.
Here, it is the neutrality of the outgoing $S$-wave pion that ensures
that the leading $\nu=-2$ terms vanish. 
The single-scattering contribution is given by the same $\nu=-1, 0, +1,...$
mechanisms described earlier, with due account of $P$-waves and 
Fermi motion inside the deuteron. 
The first two-nucleon term enters at $\nu=0$, a correction 
appears at $\nu=+1$, and so on.  
The differential cross-section at threshold, $\propto |E_d|^2$, 
was obtained at $\nu=0$ by \cite{vkolck:beane}
and up to $\nu=+1$ by \cite{vkolck:beaneetal}. 
They are shown in Table \ref{T:vkolck:Ed},
corresponding to the Argonne V18 potential and a cut-off 
$\Lambda=1000$ MeV.
Other realistic potentials and cut-offs from 650 to
1500 MeV give the same result within 5\%, 
while a model-dependent estimate 
(\cite{vkolck:wilhelm}) of some $\nu=+2$ terms 
suggests a 10\% or larger error from the neglected higher orders 
in the kernel itself. 
Some sensitivity to $E_{0+}(\pi^0 n)$ survives the large two-nucleon
contribution.  
A preliminary result from Saskatoon presented at this Workshop
is close to our prediction, while an electroproduction
experiment is under analysis at Mainz. 
For more details, see \cite{vkolck:bernard}, \cite{vkolck:bernstein},
and \cite{vkolck:merkel}.

\begin{table}[t]
\caption{Values for $E_{d}$ in units of $10^{-3}/m_{\pi^+}$
from single scattering up to $\nu=1$ (ss),
two-nucleon diagrams at $\nu=0$ (tn0), 
two-nucleon diagrams at $\nu=1$ (tn1),
and their sum (ss+tn).}
\label{T:vkolck:Ed}
\begin{tabular*}{\textwidth}{@{}l@{\extracolsep{\fill}}cccc}
\hline
ss & tn0 & tn1 & ss+tn \\
\hline
0.36 &  $-$1.90 & $-$0.25  &  $-$1.79  \\
\hline
\end{tabular*}
\end{table}

\begin{figure}[htb] 
\vspace{0.5cm}
\centerline{\epsfig{file=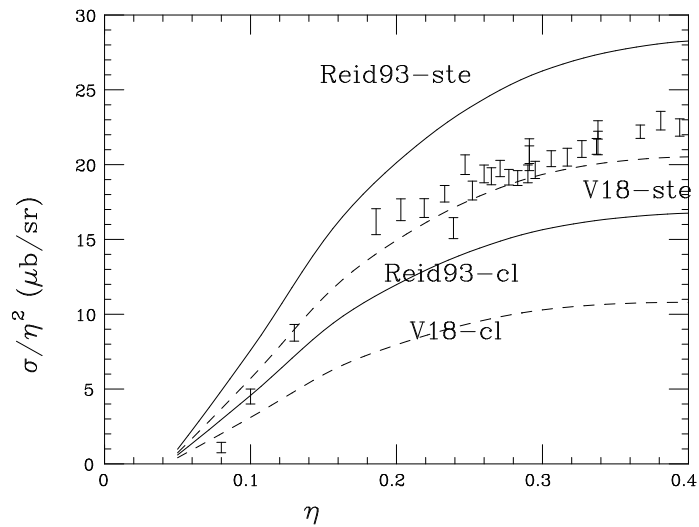,angle=270,height=2.5in,width=2.5in,
                    bbllx=1.5in,bblly=2.5in,bburx=7.5in,bbury=8.5in}}
\vspace{0cm}
\caption{Cross-section for $pp \rightarrow pp \pi^0$ in $\mu$b/sr
as function of the pion momentum $\eta$ in units of $m_\pi$ for 
two $NN$ potentials (Argonne V18 and Reid93) 
and two parametrizations of the isoscalar pion-nucleon amplitude (ste and cl).}
\label{F:vkolck:mesexfig3}
\end{figure}

{\it $\bullet \hspace{.2cm} pp \rightarrow pp\pi^0$ close to threshold.}
This reaction has attracted a lot of interest because of the failure of 
standard phenomenological mechanisms in reproducing 
the small cross-section observed near threshold. 
It involves larger momenta of $O(\sqrt{m_\pi m_N})$, so 
the relevant expansion parameter here is the not so small
$(m_\pi/m_N)^{\frac{1}{2}}$.
This process is therefore not a good testing ground for the above ideas. 
But $(m_\pi/m_N)^{\frac{1}{2}}$ is still $<1$, so at least in some formal 
sense
we can perform a low-energy expansion. 
\cite{vkolck:cohenetalpp} and \cite{vkolck:vkolcketal1} 
have adapted
the chiral expansion to this reaction and estimated the first few 
contributions. 
Again, the lowest order terms all vanish. The formally leading non-vanishing 
terms ---an 
impulse term and 
a similar diagram from the delta isobar--- 
are anomalously small and partly cancel. 
The bulk of the cross-section must then arise from contributions that 
are relatively unimportant in other processes. 
One is isoscalar pion rescattering for which two sets of $\chi$PT
parameters were used: 
``ste'' from a sub-threshold expansion of 
the $\pi N$ amplitude and ``cl'' from an one-loop analysis of threshold 
parameters. Others are two-pion exchange and short-range $\pi NNNN$ terms, 
which were modeled by heavier-meson exchange: 
pair diagrams with $\sigma$ and $\omega$ exchange, and 
a $\pi\rho\omega$ coupling, among other, smaller terms.    
Two potentials ---Argonne V18 and Reid93--- were used. 
Results are shown in Fig. \ref{F:vkolck:mesexfig3} 
together with IUCF and Uppsala data.
Other $\chi$PT studies of this reaction were carried out by
\cite{vkolck:byparketal}, \cite{vkolck:gedalin}, \cite{vkolck:sato},
and \cite{vkolck:hanhart}, 
while \cite{vkolck:park3} presented a related analysis of the 
axial-vector current. 
The situation here is clearly unsatisfactory, and presents therefore a unique 
window into the nuclear dynamics. Work is in progress, for example, on
a similar analysis for the other, not so suppressed channels
$\rightarrow d\pi^+, \rightarrow pn\pi^+$ (\cite{vkolck:carocha}). 

\section{Isospin violation}

Why is isospin a good symmetry of low-energy hadronic physics?
And charge symmetry 
(a rotation of $\pi$ around the 2-axis in isospin space)
even better?
A measure of the size of isospin violation 
compared to explicit chiral symmetry breaking in the QCD Lagrangian
is $\epsilon \sim 1/3$. 
In low-energy observables isospin violation is typically much smaller.
For example, in the $NN$ system,
using Coulomb-corrected scattering lengths and the fact that differences in 
potentials are amplified by a factor of $\sim 10$ in differences of 
scattering lengths, 
the isospin breaking but charge symmetric (or Class II) 
potential is estimated to
be around $1/40$ of the isospin symmetric (or Class I)
component, while the charge symmetry breaking (or Class III)
potential is about a further $1/4$ smaller. 

Part of the answer is of course that the quark masses
are small compared to $M_{QCD}$. This in itself does not explain, 
however, why the isospin splitting in the pion masses squared
is so small ($1/15$ of the average pion mass squared). 
It also seems to exclude the possibility of isospin violation at low
energies that is comparable to $\epsilon$.
A more complete answer requires the construction of all the operators that 
break isospin in the chiral Lagrangian.
These can be classified into three types:
{\it (i)} those operators involving hadronic fields only
that stem from the quark mass difference,
transform as 3-components of (tensor products of) 
$SO(4)$ vectors, and are proportional to (powers of) $\epsilon m_\pi^2$;
{\it (ii)} hadronic operators 
that come from exchange of hard photons among quarks,
transform as 34- and 34,34-components of 
(tensor products of) $SO(4)$ antisymmetric tensors,
and are proportional to (powers of) $e^2$;
{\it (iii)} hadronic operators that mix the above;
{\it (iv)} operators involving the photon field, which are $U(1)_{em}$
invariant and are proportional to (powers of) $e$. 
I lump type {\it (i)} operators in ${\cal L}^{qm}$,
type {\it (ii)} and {\it (iii)} in ${\cal L}^{hp}$, and
type {\it (iv)} in ${\cal L}^{sp}$.
   
One can order operators in ${\cal L}^{qm}$ using the index $\Delta$
and the power of $\epsilon$. Since a photon loop typically brings in
a power of $\alpha_{em}/\pi \sim \epsilon (m_\pi/m_\rho)^3$
one should count the index of operators in 
${\cal L}^{hp}$ (${\cal L}^{sp}$) as $\Delta+3$ ($\Delta+3/2$).
If this is done, we find (\cite{vkolck:vkolck3}, \cite{vkolck:vkolck4},
\cite{vkolck:vkolck5}) 
that isospin is an accidental 
symmetry, in the sense that its violation does not appear in the EFT
in lowest order. In most cases, an isospin violating operator 
from ${\cal L}^{qm+hp+sp}_{(n\ge 1)}$
competes
with an isospin conserving operator from ${\cal L}_{(0)}$,
so that its effects are suppressed not by $O(\epsilon)$,
but by $O(\epsilon (Q/M_{QCD})^n)$.

If one considers the pion, nucleon and delta masses one sees that the above 
naive power counting works alright. In $\pi\pi$ scattering the specific 
form of ${\cal L}^{qm}$ suggests that, when written in terms of Mandelstam 
variables, the amplitude is mostly sensitive to photon exchange effects.
In $\pi N$ scattering
we find an example of a potentially large isospin violation.
In $\pi^0$ elastic scattering,
where the isoscalar scattering length $b_0$ contributes, an
isospin violating operator from ${\cal L}^{qm}_{(1)}$
competes with an isospin conserving operator from 
${\cal L}_{(1)}$, and isospin violation could be potentially of 
$O(\epsilon)$. Unfortunately this is hard to measure directly.

In the case of nuclear forces, we recover the observed hierarchy among 
different types of components. 
In the chiral expansion, one indeed finds 
(\cite{vkolck:vkolck3}, \cite{vkolck:vkolck4}, \cite{vkolck:vkolck5},
\cite{vkolck:vkolck6}) 
that higher Class forces appear at higher orders:
$\langle V_{\rm M+1}\rangle /\langle V_{\rm M}\rangle
\sim O(Q/M_{QCD})$,
where $\langle V_{\rm M}\rangle$ denotes the contribution of 
the leading Class ${\rm M}$ potential. 
This comes about because Class I forces are dominated by
static OPE and contact terms, both with $\nu=0$,
a Class II force appears at $\nu=1$ from one insertion
of the pion mass difference in OPE, which is an 
$\epsilon m_\pi/2m_\rho \sim 1/30$ effect,
and a Class III force comes at $\nu=2$ from isospin violation 
in the $\pi NN$ coupling and in contact terms,
which is an $\epsilon (m_\pi/m_\rho)^2\sim 1/90$ effect. 
Precise calculations of simultaneous electromagnetic and strong 
isospin violation in the nuclear potential have also been carried out. 
\cite{vkolck:vkolck7} and \cite{vkolck:vkolck8}
have derived the one-pion-range isospin violating
two-nucleon potential up to $\nu=3$, which includes, besides pion and
nucleon mass splittings, also isospin violation in the pion-nucleon coupling
and simultaneous $\pi\gamma$ exchange. This is the
first calculation of this type that is both ultraviolet and infrared finite,
and independent of choice of pion field and gauge. 
This potential is comparable to other components usually considered,
and it has been included in the Nijmegen $NN$ partial-wave analysis
(\cite{vkolck:vkolck7}).

\section{Conclusions}

The challenges 
of a chiral EFT approach are greater in nuclear physics
than in the meson and  one-baryon sectors,
because of the need to go beyond a straightforward
perturbative expansion.
Despite the amount of information available,
only the very initial steps of 
a systematic chiral treatment of many-nucleon processes 
have been taken so far.

I have tried to argue that the first results are very auspicious.
The long-established picture of light nuclei as a few-body system
interacting through a non-relativistic, two-body, isospin-symmetric
potential emerges naturally at leading order. 
Higher order effects ---which account for relativistic corrections,
short-range structure and non-analytic pionic contributions---
provide the other basic ingredients of nuclear forces, 
as evidenced
by a quantitative fit to two-nucleon data and by the qualitative insights
into the size of few-nucleon and isospin-violating forces. 
$\chi$PT also provides a consistent framework for scattering on the nucleon and
on light nuclei, which in turn offers a handle on nucleon parameters
(as for pion scattering and pion photoproduction),
successful quantitative postdictions (such as in radiative neutron-proton 
capture), and
quantitative predictions (such as in pion photoproduction). 
And best of all, open problems exist 
(such as pion production in the $pp$ reaction).  
There is still a lot to be done: 
consistent potential/kernel calculations,
many other processes at and away from threshold,
extensions to $SU(3)$ and nuclear matter, to mention just a few topics.
Perhaps $\chi$PT will then fulfill the role of the long-lacking 
theory for nuclear physics based on QCD.

\vspace{.5cm}
\paragraph*{Acknowledgements:}
Section 2 benefitted from particularly helpful discussions
with P.F. Bedaque and D.B. Kaplan.
I am grateful to my collaborators for helping making this research program
possible. 
This research was supported by the DOE grant DE-FG03-97ER41014.

\end{document}